# MARTI-5: A Mathematical Model of "Self in the World" as a First Step Toward Self-Awareness

Igor Pivovarov[1] [https://orcid.org/0000-0002-5701-8717], Sergey Shumsky[2] [https://orcid.org/0009-0003-2993-0367]

[1]Moscow Institute of Physics and Technology;

[2]Symbolic Mind Inc.

Corresponding author: igorpivovarov@yandex.ru

**Abstract**

The existence of "what" and "where" pathways of information processing in the brain was proposed almost 30 years ago, but there is still a lack of a clear mathematical model that could show how these pathways work together. We propose a biologically inspired mathematical model that uses this idea to identify and separate *the self* from the environment and then build and use a self-model for better predictions. This is a model of neocortical columns governed by the basal ganglia to make predictions and choose the next action, where some columns act as "what" columns and others act as "where" columns. Based on this model, we present a reinforcement learning agent that learns purposeful behavior in a virtual environment. We evaluate the agent on the Atari games Pong and Breakout, where it successfully learns to play. We conclude that the ability to separate *the self* from the environment gives advantages to the agent and therefore such a model could appear in living organisms during evolution. We propose Self-Awareness Principle 1: the ability to separate *the self* from the world is a necessary but insufficient condition for self-awareness.

**Keywords**

Biologically inspired model; hierarchical predictive coding; reinforcement learning; symbolic agents; Atari; self-awareness; artificial intelligence; machine learning;

## 1. Introduction

In this paper, we present a mathematical model, MARTI-5, which is able to distinguish the "*self*" from the environment and effectively use this separation for purposeful behavior.

MARTI-5 forgoes neural networks and instead employs a hybrid digital-symbolic multi-agent architecture inspired by the brain. The model implements the Deep Control architecture (Shumsky, 2019) and further develops ideas proposed by Hawkins (Hawkins et al., 2016, 2017). The core idea is that the human brain should not be modelled at the level of individual neurons; rather, neurons in the neocortex form approximately $2 \times 10^8$ cortical columns (Braitenberg & Schüz, 1998; Mountcastle, 1997), each functioning as an independent module. Accordingly, MARTI-5 models the neocortex as an ensemble of cortical columns that act as micro-agents, with each column making its own predictions. All columns receive sensorimotor input from the thalamus and return their predictions to the thalamus, where these predictions are converted into motor signals directed to the cerebral cortex. The entire ensemble of cortical columns is governed by the basal ganglia, which are known to play a key role in action selection and the choice of behavior to execute (Stocco et al., 2010).

Each cortical column in MARTI-5 is implemented as a symbolic parser that constructs its own symbolic "world model". (This is, of course, a highly simplified representation of a biological cortical column; however, it is sufficient to capture the core concept of a predictive agent while preserving transparency, human interpretability, and computational efficiency.)

Drawing on the classic distinction between the dorsal ("where") and ventral ("what") pathways of visual information processing (Ungerleider & Mishkin, 1982; Goodale & Milner, 1992), the cortical columns in MARTI-5 are divided into two functional groups: "what" columns and "where" columns.

"What" columns receive complete information about the current state of the environment, including the agent's own state (analogous to viewing the world and seeing oneself as part of it). In contrast, "where" columns receive only partial information – specifically, the subset of input features hypothesized to correspond to the agent's own state.

The model begins from scratch with no prior knowledge of how the "*self*" is represented in the input data. The sole assumption is that such a representation exists and that changes in the agent's own state occur at a rate comparable to changes in the overall sensory input.

The core principle for distinguishing the self from the environment is based on the observation (Apps & Tsakiris, 2014; Allen & Tsakiris, 2019) that one's own body is the most predictable – and therefore the most controllable – part of the observable world. MARTI-5 exploits this principle by analysing the predictability and controllability of different components of the sensory input.

To achieve this, the model initially selects several highly active input features and treats each of them as a candidate for representing the agent itself (the exact selection process is detailed in Section 3). For every candidate, MARTI-5 creates a corresponding set of "where" columns and evaluates whether that component is significantly more predictable and controllable than the environment on average.

Each "where" column receives, in addition to its input feature, information about the agent's most recent action. It then computes the pointwise mutual information (PMI) coefficient, which quantifies how much knowledge of the previous action increases the predictability of the next state of that feature. Formally, for a given "where" feature $w_n$:

$$PMI = \log_2 [ P(w_n | a_{n-1}, w_{n-1}) / P(w_n | w_{n-1}) ]$$

where $a_{n-1}$ is the previous action. If the average PMI exceeds a predefined threshold (typically PMI > 1), it indicates that the previous action strongly influences the next state of this feature, meaning the feature is highly controllable. Such a column is then confirmed as a genuine "where" column representing part of the agent's *self*.

For example, in the Pong game, the position of the agent's own racket is strongly influenced by the agent's previous actions, whereas the position of the opponent's racket is largely independent of those actions. Consequently, a "where" column analysing the input feature corresponding to the "self" racket will be able to predict its next state far more accurately when conditioned on the previous action; it will therefore exceed the PMI threshold and be confirmed. In contrast, a "where" column tied to the opponent's racket will show little or no improvement in predictability and will fail the confirmation test.

Most candidate "where" columns are eventually rejected, and the corresponding structures are pruned. This mechanism initially generates a deliberately redundant representation of the sensory input, which is subsequently refined and reduced – a process strikingly similar to synaptic overproduction followed by competitive pruning observed in early brain development (Changeux et al., 1973; Rakic et al., 1986; Innocenti & Price, 2005).

Once the confirmed "where" columns have been identified, MARTI-5 can exploit them for prediction and action selection. We propose a simple yet effective interaction mechanism between "what" and "where" columns that explicitly uses the discovered self-representation to improve action prediction.

The core prediction process works as follows:

First, the ensemble of "what" columns jointly identifies the most desirable next global state (the "what" – i.e., the optimal situation the agent should reach).

Then, using the confirmed self-representation, the "where" columns determine which action is most likely to move the agent's own state from its current location $A$ to the target location $B$ that corresponds to this desired global state (the "where" – i.e., the motor command required).

More precisely, each confirmed "where" column receives the current self-state $A$ and the target self-state $B$ (extracted from the best predicted global state) and proposes the action that maximises the transition probability $P(B|A, \text{action})$. The final action is selected via majority voting across the ensemble (see Section 3.4).

Importantly, prediction is not performed by a single unified model but emerges from the parallel, distributed predictions of multiple independent columns. At each decision step, $K$ "what" columns simultaneously forecast the most valuable next global situation, and $K$ confirmed "where" columns simultaneously infer the action most likely to achieve the corresponding self-state transition.

The final action is selected through a majority vote across the ensemble of confirmed "where" columns, with the striatum (modelled here as part of the basal ganglia) orchestrating the process. In biological organisms, the basal ganglia evaluate candidate actions by integrating a broad range of internal states and drives – including hunger, fear, energy levels, curiosity – alongside cortical input; the cortical ensemble is thus only one instrument in a much larger orchestra. In the simplified environments used in the current work, however, these additional motivational states are absent. Consequently, the model's striatum considers only the votes from the cortical columns and selects the action supported by the majority of confirmed "where" columns (or chooses a random action if no clear majority emerges; see Section 3.4).

When the agent receives an external reward or penalty (typically at the end of an episode), this scalar signal is broadcast to all "what" parsers. Each parser updates a value function over the symbols in its vocabulary, enabling it to estimate which successor states are most desirable in the future.

To complete the overview, MARTI-5 is designed to support hierarchical predictive coding in principle. However, in the experiments presented here we deliberately employ a flat two-layer architecture without higher-level temporal abstraction or hierarchical planning, allowing us to isolate and study the core mechanisms of self-model discovery and its immediate exploitation for action selection.

## 2. Related Work

The notion that self-awareness arises from the fact that one's own body is the most predictable and controllable part of the sensorium represents a synthesis of ideas spanning more than 150 years – from Hermann von Helmholtz's 19th-century theories of perception-as-inference to contemporary computational neuroscience.

An early formal approach to self-referential processing in agents was proposed by Polani and colleagues through the concept of empowerment (Klyubin et al., 2005). Empowerment – defined as the channel capacity of the sensorimotor loop – provided a crucial proof-of-concept that a single information-theoretic quantity can give rise to sophisticated survival- and exploration-oriented behavior without explicit reward functions.

Subsequently, Karl Friston introduced a unifying formalisation via the Free Energy Principle (Friston et al., 2006; Friston, 2009, 2010), describing living systems as active inference agents that strive to minimise future surprise (variational free energy). This framework has proved profoundly influential in establishing a rigorous scientific foundation for understanding adaptive behaviour. Key contributions by Rao (Huang & Rao, 2011), Clark (2013, 2015), Hohwy (2016), and Seth (Seth & Tsakiris, 2018; Seth, 2015) have progressively extended these ideas into a comprehensive predictive processing account that explicitly links prediction error minimisation, interoception, and the emergence of self-consciousness.

Modern experimental evidence – ranging from studies of interoceptive accuracy (Seth et al., 2012) to fMRI investigations of self-referential processing (Moran et al., 2013) – increasingly supports the view that self-awareness arises from the brain's exceptionally precise and continuously updated predictions about its own embodied states.

Deep reinforcement learning (DRL) approaches to mastering Atari games have become a cornerstone of modern AI agents research. This line of work began with the landmark Deep Q-Network (DQN) (Mnih et al., 2013) and has since seen major advances, including Rainbow (Hessel et al., 2018) and Agent57 (Badia et al., 2020).

The core biological mechanisms implemented in the MARTI-5 model were originally proposed by Hawkins and colleagues (Hawkins et al., 2017). The concept of hierarchical predictive coding in the neocortex, which MARTI-5 also adopts, was introduced by Friston (2005) and further elaborated by Bastos et al. (2012), Clark (2015), and Spratling (2017). Unlike earlier neocortical models (Hawkins & Subutai, 2016; Laukien et al., 2016), the Deep Control architecture uniquely integrates Hebbian-style predictive learning in the cortex with reinforcement learning in the basal ganglia, thereby realising the so-called "super-learning" hypothesis (Caligiore et al., 2019).

## 3. The model

MARTI-5 receives as input a digital vector that fully describes the current state of the environment (including the agent's own state) along with a scalar reinforcement signal: +1 for a won episode and –1 for a lost episode. The model outputs a discrete action (e.g., move left, move right, or stay). Its objective is to maximise cumulative reward.

In contrast to the earlier MARTI-4 model (Pivovarov & Shumsky, 2022), which could only exhibit learned behavior, MARTI-5 is capable of discovering a representation of "*self*" within the sensory input and separating it from the rest of the "world". This ability dramatically improves both learning efficiency and final performance, while also making the decision-making process far more transparent and interpretable. In the present paper we do not describe every technical detail of the architecture; instead, we focus on the mechanisms of self-model discovery and its subsequent exploitation.

MARTI-5 begins from scratch with only a minimal set of developmental rules. The model grows incrementally "on the fly": its size and complexity are determined solely by the quantity and diversity of the experienced data. This stands in sharp contrast to deep neural networks (DNNs), which are pre-allocated with billions of parameters from the outset and require repeated forward and backward passes over the entire network during training.

The model consists of several key components:

### 3.1 Thalamus

In the human brain, the thalamus is thought to act as an "information hub", relaying sensorimotor input from sensory organs to neocortical columns and routing the columns' outputs back to motor neurons.

In MARTI-5, the thalamus is implemented as a single object. It receives a digital input vector from the environment, orchestrates the entire process of information processing and prediction, and returns the model's chosen action to the environment.

The input vector consists of three parts:

- a full state vector representing the current environment (including agent's own state);
- a vector encoding the previous action;
- a scalar reward (or penalty).

The thalamus creates a single "what" connector that receives the full state vector. At the beginning, the model has no prior knowledge of which components of this vector represent the agent's own state. To discover them, the thalamus first identifies the 10% most actively changing coordinates in the full vector and treats each of these coordinates as a candidate for the agent's state representation. For each such candidate, the thalamus creates a separate "where" connector. (Most of these "where" connectors will later fail the confirmation test and be discarded.)

The thalamus processes the incoming vector as follows:

- The full environment state vector, together with the current reward, is passed to the "what" connector.
- Each candidate "agent-state" coordinate is passed to its corresponding "where" connector.
- The vector encoding the previous action is passed to the "action" connector.

### 3.2 Connector

Connectors are multiple objects in the model, organized in layers. Each connector is created by the thalamus and receives digital vectors as input. After accumulating a sufficient number of vectors, the connector performs **K** independent clusterings of the collected data. Thereafter, for any new input vector, it can assign a cluster index (0, 1, …) independently for each of the **K** clusterings. These cluster indices serve as symbolic tokens (0 = **A**, 1 = **B**, etc.) that are fed to the downstream parsers.

There are three types of connectors, distinguished by the input they receive:

- The "where" connector receives the full state vector of the environment (including the agent's own state). Consequently, each cluster in any of its **K** clusterings represents a set of similar global environment states.
- The "what" connector receives only a single coordinate from the full state vector – the one that potentially represents the agent's own state. Consequently, each cluster corresponds to a set of similar values of this particular coordinate..
- The "action" connector receives the vector that encodes the previously executed action. Consequently, each cluster represents one specific action available to the agent.

### 3.3 Parsers

The parser represents the basic computational unit that models a single cortical column. Each parser builds its own symbolic "world model" from the stream of incoming symbols and uses it to predict the next state. The overall predictive power of MARTI-5 is strongly correlated with the total number of active parsers (i.e., cortical columns).

Parsers are multiple objects in the model, organized in layers. Each parser is instantiated on top of one specific clustering (one of the **K** independent clusterings) of a single connector and receives the corresponding symbolic token (cluster index) as its input.

There are two types of parsers, determined by the type of their parent connector:

#### 3.3.1 Parser "what"

This parser receives symbolic tokens from its parent "what" connector. It processes the incoming stream of symbols, discovers sequential patterns and regularities, and builds a predictive world model.

To this end, the parser maintains:

- a vocabulary **S** containing all symbols encountered so far;
- a correlation table **C** that records co-occurrence statistics between consecutive symbols.

Each time the parser receives a new symbol $s_n$, the correlation table is updated as follows:

$s_n \rightarrow C(s_{n-1}, s_n) \leftarrow C(s_{n-1}, s_n) + 1$

Whenever the vocabulary already contains **m** symbols and the correlation between two consecutive symbols $s_{n-1}$ and $s_n$ exceeds a predefined threshold **T**, the parser creates a new composite symbol (a "word") provided that its length does not exceed the maximum allowed word size **W**. This new symbol is then added to the vocabulary:

if $C(s_{n-1}, s_n) > T \rightarrow s_{m+1} := s_{n-1}s_n$

(Here the concatenation $s_{n-1}s_n$ denotes a new higher-order symbol representing the bigram.)

The parser's predictions are driven by a learned value function. To support this, it maintains a reward table **R** that stores cumulative rewards associated with symbol transitions (i.e., $R(s_i, s_j)$ represents the expected reward following the transition $s_i \rightarrow s_j$).

At every time step the parser receives a reward signal: either the external environment reward or, when no external reward is available, a small internal step penalty of **−0.1**. External rewards (game win = +1 or game loss = −1) are broadcast to all "what" parsers simultaneously.

When an external reward $r_n$ is received at step **n**, the parser updates not only the most recent transition $s_{n-1} \rightarrow s_n$, but also the preceding **M** transitions, using an exponentially decaying memory factor **k** (0 < k < 1). Updates are performed with a fixed learning rate **α** (rather than the unstable **1/n** schedule):

$r_n \rightarrow$

$R(s_{n-1}, s_n) \leftarrow (1-\alpha) R(s_{n-1}, s_n) + \alpha\, r_n$
$R(s_{n-2}, s_{n-1}) \leftarrow (1-\alpha) R(s_{n-2}, s_{n-1}) + \alpha\, k\, r_n$
$R(s_{n-3}, s_{n-2}) \leftarrow (1-\alpha) R(s_{n-3}, s_{n-2}) + \alpha\, k^2\, r_n$
⋮

Thanks to these accumulated values, the parser can predict the next symbol $s_{n+1}$ that maximizes expected future reward:

$s_{n+1} = \mathrm{argmax}_i\, R(s_n, s_i)$

### 3.3.2 Parser "where"

This parser receives two symbolic tokens simultaneously: one from its parent "where" connector (representing the candidate agent-state coordinate) and one from the "action" connector. It constructs its input symbol as an ordered pair ⟨$a_{n-1}$, $w_n$⟩ (action–where). Thus, every internal symbol $s_n$ effectively represents the joint event "previous action $a_{n-1}$ followed by the current candidate self-state $w_n$".

Like the "what" parser, it processes the stream of such paired symbols and discovers sequential regularities. To this end, it maintains:

- a vocabulary **S** of all observed pairs, and
- a correlation table **C** that counts co-occurrences of consecutive pairs.

Each time the parser receives a new paired symbol $s_n$ = ⟨$a_{n-1}$, $w_n$⟩, the table is updated as:

$$s_n \rightarrow C(s_{n-1}, s_n) \leftarrow C(s_{n-1}, s_n) + 1$$

Unlike the "what" parser, the "where" parser does not create higher-order composite symbols (words).

Its prediction is purely statistical and chooses the successor symbol $s_i$ that has been observed most frequently after the current $s_n$:

$$s_{n+1} = \mathrm{argmax}_i\, C(s_n, s_i)$$

(This is equivalent to maximum-likelihood prediction of the next joint pair ⟨$a_n$, $w_{n+1}$⟩ given the recent history.)

Each "where" parser operates in one of two states: **potential** or **confirmed**. While in the **potential** state, the parser continuously evaluates whether knowledge of the previous action significantly improves the predictability of the next value of its candidate coordinate. To quantify this, it computes the pointwise mutual information (PMI) between the action-conditioned and unconditional transitions:

- **P($w_n$)** = estimated probability of the next self-coordinate value $w_n$ without knowing the previous action
- **P($w_n$ | $w_{n-1}$)** = probability of $w_n$ given only the previous value of the same coordinate
- **P($s_n$ | $s_{n-1}$)** = probability of the full paired symbol $s_n$ = ⟨$a_{n-1}$, $w_n$⟩ given the previous paired symbol $s_{n-1}$ (i.e., conditioned on both previous action and previous coordinate value)

The pointwise mutual information is then calculated as:

$$\mathrm{PMI} = \log_2 [\, P(s_n | s_{n-1}) / P(w_n | w_{n-1})\, ]$$

This value measures (in bits) how much more predictable the next coordinate value becomes when the previous action is taken into account.

After **N** prediction steps (typically several thousand), the parser examines its average PMI:

- If **PMI ≳ 1 bit** (i.e., knowing the action at least doubles the probability of correct prediction on average), the parser is marked as **confirmed** – the corresponding coordinate is considered part of the true self-representation.
- Otherwise (PMI ≈ 0 or negative), the parser is marked as **failed** and is subsequently destroyed along with its connector.

Confirmed "where" parsers are then paired with corresponding "what" parsers for joint prediction and action selection (see Section 3.5).

### 3.4 Striatum (Basal Ganglia)

In the biological brain, the striatum is the principal input structure of the basal ganglia and receives massive convergent projections from virtually all neocortical areas. It integrates cortical predictions with internal motivational signals (hunger, fear, curiosity, etc.) to select the most appropriate action at each moment.

In MARTI-5, this functionality is implemented by a single striatum object that collects predictions from all active parsers in every layer and orchestrates action selection via majority voting among confirmed "where" parsers. Specifically:

- Each confirmed "where" parser votes for the action that, according to its own statistics, is most likely to produce the desired next self-state.
- The striatum selects the action that receives the absolute majority of votes.
- If no action gathers at least **T** votes (a configurable confidence threshold, i.e., when the ensemble is uncertain), the striatum falls back to a uniformly random action.

This simple yet biologically plausible voting mechanism serves as the final decision stage of the model.

### 3.5 Structure of the Mature Model

Figure 1 presents a schematic of the fully developed MARTI-5 architecture after approximately 5000 training games.

*Figure 1. Mature MARTI-5 model after 5000 games of training (K = 25 clusterings). Connector 1 is the single "what" connector receiving the full environment state. Connector 2 is the sole surviving (confirmed) "where" connector that successfully identified the true agent-state coordinate. Parsers 1–25 labelled "ParserSelfAction" are the confirmed "where" parsers, while Parsers 1–25 labelled "ParserAction" are the corresponding*

"what" parsers. All failed "where" connectors and their associated parsers have been pruned.

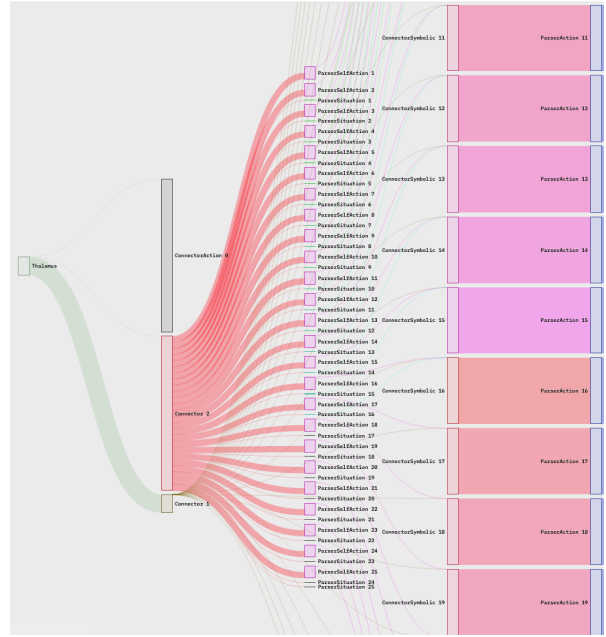

### 3.6 Overall Processing Pipeline

The complete forward pass of MARTI-5 consists of two alternating phases executed every time step: the **IN** (perception) phase and, every k-th step, the **OUT** (prediction and action-selection) phase.

*IN phase (executed at every time step)*

When the model receives a new input vector from the environment, the thalamus initiates the IN propagation:

1. The full environment state vector together with the current reward is forwarded to the "what" connector.
2. Each candidate agent-state coordinate is forwarded to its corresponding (potential) "where" connector.
3. The vector encoding the previous action is forwarded to the "action" connector.

Then, layer by layer, the thalamus propagates the information downward:

- Each connector determines the cluster index that best matches its current input (separately for each of its **K** independent clusterings) and passes the resulting symbolic token(s) to the child parsers.
- Each parser updates its internal state according to the received symbol (incrementing the relevant entry in its correlation table **C** and, for "what" parsers, processing the reward signal as described in Section 3.3.1).

*OUT phase (executed every k-th time step – frame-skipping)*

Following common practice in Atari reinforcement learning (Bellemare et al., 2013; Mnih et al., 2013), the model repeats the last selected action on skipped frames and performs full prediction only every **k** steps (k = 3 in our experiments). When an OUT phase is triggered:

1. Starting from the deepest layer and moving upward, every parser generates its prediction of the next symbol according to its own statistics (value-based for "what" parsers, frequency-based for confirmed "where" parsers).
2. Predictions are propagated back through the corresponding connectors to lower layers when needed.
3. All confirmed "where" parsers independently vote for the action they consider most likely to reach the globally preferred next situation (as determined by the ensemble of "what" parsers).
4. The striatum collects the votes and selects the action with the absolute majority. If no action receives at least **T** votes, a random action is chosen instead.
5. The selected action is sent back to the environment via the thalamus.

This interleaved IN/OUT scheme ensures continuous online learning while keeping computational cost low.

## 4. Experimental Setup

We evaluated MARTI-5 on two classic Atari 2600 games: Pong and Breakout, using the Gymnasium environment (the current version of the original OpenAI Gym; Brockman et al., 2016). Both games present a challenging sparse-reward setting in which non-zero rewards (±1) are delivered only at the end of each episode.

The Atari Learning Environment offers two observation modes: raw screen images (84×84 or 210×160 pixels) and the 128-byte RAM state. Although the RAM state is fully informative (Bellemare et al., 2013), deep RL agents typically achieve better results when learning from raw screen images, which retain explicit spatial relationships that are only implicitly encoded in the RAM. We nevertheless opted for RAM observations in the present work, as they better align with our hypothesis that neocortical processing operates on already-preprocessed sensory representations (analogous to the output of early visual areas). This choice also dramatically reduces input dimensionality and computational cost. Future extensions will incorporate convolutional preprocessing to handle raw pixels directly.

The agent received no game-specific information, no hand-crafted features, and no access to the emulator's internal variables. It observed only the 128-byte RAM vector, the current reward, episode termination signals, and the set of available discrete actions.

All experiments were conducted on a standalone single-CPU server (AMD Ryzen 9 3900X, 24 threads, 128 GB RAM) running Ubuntu 22.04. The MARTI-5 core is implemented in Java (OpenJDK 17.0.11) and communicates with Gymnasium via a lightweight TCP/IP socket interface. A small Python wrapper script converts Gymnasium RAM observations and rewards into text format, forwards them to the Java process, and returns the selected action back to Gym.

Under these conditions, the full model processes one environment step in approximately 2 ms on average, enabling roughly 50,000 complete games (≈170 million environment steps) to be run in about 100 hours of wall-clock time.

Training episodes were capped as follows:

- Pong: maximum 18,000 frames or until one player reaches 21 points.
- Breakout: maximum 18,000 frames or until the agent loses its 5 lives.

## 5. Results

The results reported below were obtained with the following hyper-parameters:

- Number of independent clusterings $K = 50$
- Total "what" clusters = 1600
- Total "where" clusters per connector = 75
- Learning rate $\alpha = 0.3$
- Memory decay factor $k = 0.9$
- Reward propagation depth $M = 20$
- Maximum word length $W = 3$
- Frame-skipping period $k = 3$ (action repeated every 3 frames)
- Minimum votes for confident action selection $T = 3$

MARTI-5 successfully learns competent policies for both Pong and Breakout, transitioning from initially random behaviour to stable, goal-directed control within the 50,000 training games.

**Table 1** summarizes the final performance after 50,000 games of online training, alongside several relevant baselines.

*Table 1. Performance of MARTI-5 after 50,000 training games compared with published baselines. Reported values are average over the last 500 evaluation games (where available) or best reported scores for the baselines.*

|  | Average difference between won and lost points | |
|---|---|---|
|  | Pong | Breakout |
| Random | -20.4 | -4 |
| Human | -3 | 31 |
| SARSA (Bellemare et al., 2013) | -19 | 5.2 |
| MARTI-4 (previous version) | -15,8 |  |
| **MARTI-5 (this work)** | **-11.5** | **50** |
| DQN (Mnih et al., 2015) | 20 | 168 |
| **MARTI-5 – best game** | **9** | **420** |
| DQN – best game | 21 | 225 |

MARTI-5 substantially outperforms random play, the tabular SARSA baseline, average human performance, and our earlier MARTI-4 architecture on both games. Although its final scores remain below the classic DQN benchmark, it is worth noting that (i) MARTI-5 learns continuously online without experience replay or separate training/inference phases, and (ii) it achieves these results on a single CPU in real time.

It is important to highlight a fundamental difference in the learning paradigm between deep neural networks and MARTI-5.

Classic deep RL agents, such as DQN (Mnih et al., 2013), separate learning into two distinct phases: an offline training stage that processes a fixed replay buffer of approximately 10 million frames over many epochs (effectively exposing the network to roughly 1 billion environment steps), followed by a pure inference stage in which no further weight updates occur.

In contrast, MARTI-5 has no separation between training and inference. The model is instantiated from scratch and learns continuously online, without experience replay or multiple passes over the same data. The results reported here were obtained after 50,000 complete games (approximately 170 million environment steps), executed in real time on a single CPU in approximately 100 hours of wall-clock time.

Given the radically different learning dynamics of this symbolic, continuously growing architecture, it is plausible that optimal performance requires substantially longer training than the 50,000 games presented here. Extending training to hundreds of thousands or millions of games is a natural direction for future work, pending greater computational resources.

**Figure 2** shows a typical learning curve for Breakout over 50,000 training games. The characteristic saw-tooth pattern and high variance arise from the game's scoring system: points per brick range from 1 to 7 depending on the row, so a single successful game can yield anywhere between approximately 60 and 420 points. Consequently, the moving average fluctuates considerably until the policy stabilizes.

*Figure 2. Learning curve for MARTI-5 on Breakout (50,000 games). Green dots: running average over the last 25 games; black dots: running average over the last 500 games.*

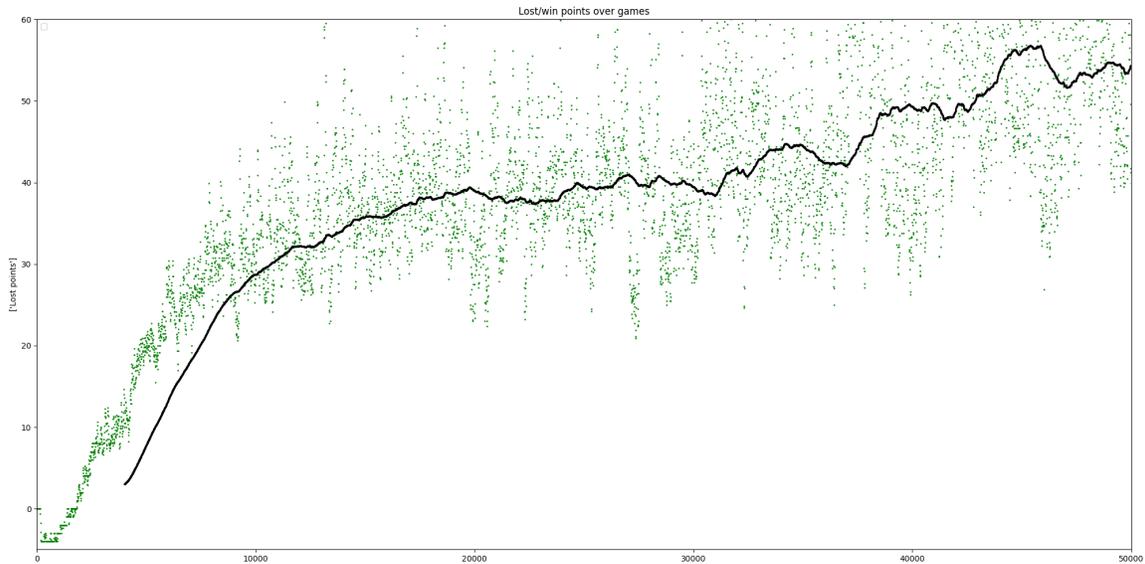

Video 1. MARTI-5 gameplay on Breakout after 50,000 training games (or click here to view: https://www.dropbox.com/scl/fi/a5iv1wihv7qv19gp3lplc/Marti-5.2_breakout.gif?rlkey=d77m3oia6utjeccej76f48imb&dl=0 )

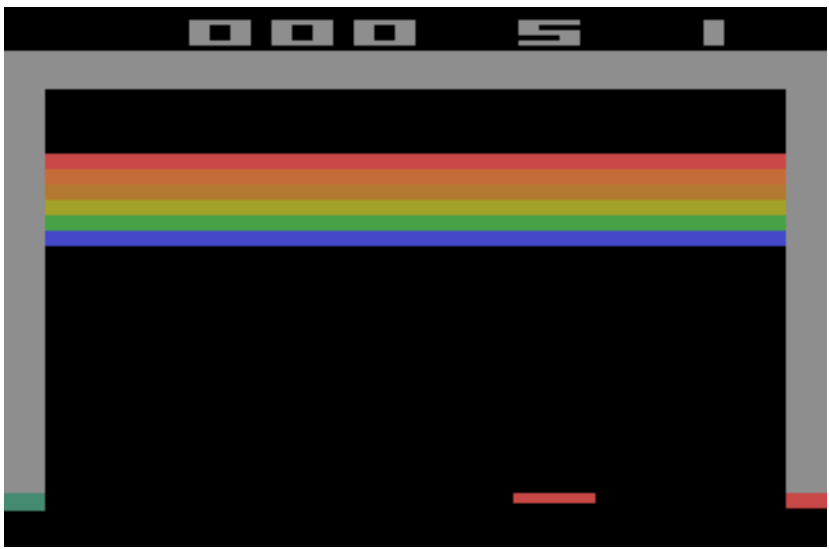

**Video 2.** MARTI-5 gameplay on Pong after 20,000 training games. (or click here to view: https://www.dropbox.com/scl/fi/d59npdvbsuwoxbvzevxb0/Pong_trained.mp4?rlkey=4bayzqw1wjq084v7j4ww7txyn&dl=0)

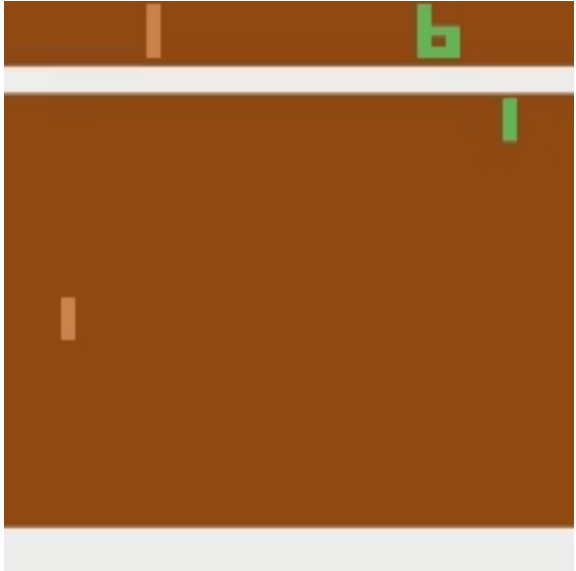

## 6. Discussion

Following Anokhin (2021), phenomenal consciousness can be decomposed into several hierarchical levels: sensory, affective, cognitive, agentive, and self-awareness.

Deep neural networks already implement credible analogues of the first three levels. They receive sensory input (images, text, or other modalities), perform sophisticated internal computations that can be interpreted as cognitive processing, and – when embedded in a reinforcement-learning loop – exhibit goal-directed agency. Thus, modern DNN-based agents possess clear models of sensory, cognitive, and agentive experience.

However, no current DNN architecture contains an explicit, identifiable mechanism corresponding to self-awareness. Claims that today's large language models or vision models are self-aware, therefore, remain largely speculative and lack empirical grounding.

In contrast, MARTI-5 incorporates a biologically inspired, computationally explicit procedure whose sole purpose is to distinguish the controllable and highly predictable subset of the sensory field (the agent's own state) from the rest of the environment. This separation constitutes the first necessary step toward genuine self-representation.

We do not claim that MARTI-5 possesses phenomenal self-awareness – its "self-model" remains purely functional and symbolic. Nevertheless, the very fact that a relatively simple architectural modification (splitting processing into "what" and "where" pathways plus a predictability-based confirmation loop) yields both (i) a measurable performance advantage and (ii) an identifiable self-representation supports the hypothesis that the evolutionary pressure to separate self from world may have been a key driver in the emergence of cortical architecture and, eventually, of self-awareness itself.

Thus, the ability to delineate the boundary between self and non-self appears not only computationally beneficial (as demonstrated by the improved Atari scores) but also conceptually foundational: it is a necessary (albeit insufficient) precondition for any higher form of self-awareness.

We expect a natural objection, that selecting a few highly controllable coordinates from the input vector is mere "agent localization" rather than genuine self/non-self distinction.

This objection would carry significant weight in richer, embodied environments (e.g., a robot with dozens of proprioceptive channels, vestibular input, pain, touch, and interoceptive signals). In everyday human experience, the sense of self indeed encompasses an entire subjective inner world – body ownership, affective tone, narrative identity, and more.

However, the Atari domain is deliberately minimal: the agent is effectively one-dimensional (a paddle or a ball-moving hand), and its "body" consists of just a few RAM bits that change predictably with its own actions. In this extremely reduced setting, localizing the agent's own controllable degrees of freedom within the observation vector is functionally equivalent to constructing a self-representation. There is nothing else in the input stream that could plausibly serve as "the self."

Thus, the criticism reduces to a mismatch of complexity rather than principle: MARTI-5's mechanism is correctly solving the self-detection problem that the Atari environment actually poses. When the same architectural principle is later applied to far richer sensory streams (multi-modal, high-dimensional, truly embodied), the resulting self-model will scale accordingly – and the distinction between "mere localization" and "true self-representation" will largely dissolve.

A second anticipated objection is that deep neural networks might implicitly develop something functionally similar to the explicit self-model described here, perhaps in the form of internal representations that systematically track the agent's own state across time, even if this was never an architectural requirement.

This possibility cannot be ruled out, and we are not claiming that DNNs are incapable of any form of self-representation. However, three crucial differences remain:

- **Explicit vs. implicit**: In MARTI-5 the self-model is architecturally explicit, interpretable, and arises from a dedicated, biologically motivated detection mechanism. In contrast, any analogous representation in a DNN would be an emergent by-product buried in distributed activations and extremely difficult to identify or interpret.
- **Necessity of self-in-the-world representation**: Building a usable self-model requires that the agent's own state be observable within the sensory input stream. This condition is satisfied in embodied RL settings (Atari, robotics, etc.), but is largely absent in the training data of most large language models and many vision models, where the system has no consistent, controllable "body" signal. Without such a signal, no meaningful self/non-self separation can emerge – even implicitly.
- **Architectural pressure**: MARTI-5 actively exploits the detected self-boundary to improve prediction and control. In standard DNNs there is no comparable architectural incentive or dedicated circuitry to maintain and refine such a boundary.

Thus, while deep networks may incidentally learn useful internal regularities that partially resemble a self-model, the deliberate, transparent, and theoretically motivated separation implemented in MARTI-5 represents a qualitatively different approach – one that more closely mirrors proposed neurobiological mechanisms of self-awareness.

Finally, returning to the classic dissociation between the ventral ("what") and dorsal ("where/how") streams in the mammalian visual cortex, we propose that their evolutionary origin and persistent specialization may be largely explained by the computational necessity of constructing and maintaining an explicit self-model. The ventral stream provides a rich, action-independent description of the world, whereas the dorsal stream is dominated by fast, egocentric, action-contingent signals – precisely the contrast required for reliable separation of the most controllable and predictable subset of the sensory field (the bodily self) from the rest of the environment. In this view, the primary evolutionary purpose of the two-stream architecture is not merely object recognition versus spatial navigation, but the continuous, predictive delineation of self versus non-self to enable more accurate forecasting and adaptive control.

## 7. Conclusions

We have presented MARTI-5: a biologically inspired, computationally explicit model of neocortical–basal ganglia–thalamic interaction. The architecture represents the neocortex as a large ensemble of cortical columns operating as semi-independent predictive micro-agents. Action selection is governed by a striatum-like voting mechanism that integrates predictions across the entire ensemble.

A core contribution is a novel, biologically plausible mechanism by which "what" (ventral-like) and "where" (dorsal-like) columns interact to automatically detect the agent's own controllable state within the sensory input stream. This detection relies solely on predictability and action-contingency criteria – no prior knowledge of the input structure is required. The resulting self-model is then actively exploited to improve next-state prediction and action selection.

Using only raw Atari RAM observations and sparse end-of-game rewards, MARTI-5 successfully learns competent policies for Pong and Breakout from scratch on a single CPU, substantially outperforming random play, tabular baselines, average human performance, and our previous MARTI-4 architecture.

These results provide empirical support for the hypothesis that the ability to separate self from non-self confers a significant selective advantage in predictive control. We therefore suggest that the evolutionary emergence of dedicated neuroarchitectural mechanisms for self/non-self segregation (exemplified by the two visual streams) may have been driven primarily by this computational benefit, and that such segregation constitutes the foundational, necessary (though not sufficient) step toward genuine self-awareness.

Accordingly, we formulate:

**Self-Awareness Principle 1** The ability to distinguish *the self* from the world is a necessary, but not sufficient, condition for self-awareness.

**Corollary 1** Self-awareness cannot emerge in any system that lacks a mechanism for delineating the boundary between its own controllable states and the external environment.

Future work will scale the same architectural principles to richer, multimodal, and truly embodied environments, where the resulting self-models are expected to become correspondingly more sophisticated and phenomenologically relevant.

**Declaration of generative AI and AI-assisted technologies in the manuscript preparation process.**

During the preparation of this work authors used Grok 4.1 in order to check the grammar and rewrite particular pieces of final text for better readability. After using this service, authors reviewed and edited the content as needed and take full responsibility for the content of the published article.

**Declaration of competing interest**

The authors declare that they have no known competing financial interests or personal relationships that could have appeared to influence the work reported in this paper.

**Funding**

This research did not receive any specific grant from funding agencies in the public, commercial, or not-for-profit sectors. The work was made by authors solely on their own time without any support or funding from any organization.

## References

Allen M. and Manos T., 2019. The body as first prior: Interoceptive predictive processing and the primacy of self-models.

Allen M. and Tsakiris M, 2019. The body as first prior: Interoceptive predictive processing and the primacy. In The Interoceptive Mind: From Homeostasis to Awareness, 1st ed. (pp. 27–45). Oxford University Press.

Anokhin K.V., 2021. Cognitom: in a search for fundamental scientific neurophysiological theory of consciousness. I.P. Pavlov Journal of higher nervous activity. V. 71. № 1. pp. 39–71.

Apps M.A., and Tsakiris M., 2014. The free-energy self: a predictive coding account of self-recognition. Neuroscience & Biobehavioral Reviews 41: 85-97. doi: 10.1016/j.neubiorev.2013.01.029

Bacon P.L, Jean H. Doina P., 2017 The option-critic architecture. Proceedings of the AAAI Conference on Artificial Intelligence. Vol. 31. No. 1.

Badia A.P., Piot B., Kapturowski S., Sprechmann P., Vitvitskyi A., Guo Z.D., Blundell C., 2020. Agent57: Outperforming the atari human benchmark. International conference on machine learning 2020 Nov 21 (pp. 507-517). PMLR.

Bastos A.M., Usrey W.M., Adams R.A., Mangun G.R., Fries P., Friston K.J., 2012. Canonical microcircuits for predictive coding. Neuron. Nov 21;76(4):695-711. doi: 10.1016/j.neuron.2012.10.038.

Bellemare G. Marc et al., 2013. The Arcade Learning Environment: An Evaluation Platform for General Agents, Journal of Artificial Intelligence Research 47, 253-279

Botvinick M.M., 2012. Hierarchical reinforcement learning and decision making. Current opinion in neurobiology 22.6: 956-962.

Braitenberg V. and Schüz A., 1998. Cortex: Statistics and Geometry of Neuronal Connectivity. Springer, Berlin. doi:10.1007/978-3-662-03733-1

Brockman G. et al., 2016. Open AI Gym. arXiv

Caligiore D., Arbib M.A., Miall R.C., Baldassarre G., 2019. The super-learning hypothesis: Integrating learning processes across cortex, cerebellum and basal ganglia. Neuroscience & Biobehavioral Re-views 100: 19-34.

Changeux, J.P., Courrbge, P., Danchin, A., 1973. A theory of the epigenesis of neural networks by selective stabilization of synapses. Proceedings of the National Academy of Science USA, 70,2974-2978.

Clark A., 2013. Whatever next? Predictive brains, situated agents, and the future of cognitive science. Behavioral and brain sciences. Jun;36(3):181-204.

Clark A., 2015. Surfing uncertainty: Prediction, action, and the embodied mind. Oxford University Press.

Friston K.J., 2005. A theory of cortical responses. Philosophical transactions of the Royal Society B: Biological sciences 360.1456: 815-836.

Friston K.J., Kilner J., Harrison L., 2006. A free energy principle for the brain. Journal of Physiology-Paris. 100 (1–3): 70–87. doi:10.1016/j.jphysparis.2006.10.001

Friston K.J., 2009. The free-energy principle: A rough guide to the brain?, Trends in Cognitive Sciences, 13 (7), pp. 293–301.

Friston, K.J., 2010. The free-energy principle: a unified brain theory? Nature Reviews Neuroscience 11, 127-138

Goodale M.A., Milner A.D., 1992. Separate visual pathways for perception and action. Trends Neurosci. Jan;15(1):20-5. doi: 10.1016/0166-2236(92)90344-8. PMID: 1374953.

Hawkins J. and Subutai A., 2016 Why neurons have thousands of synapses, a theory of sequence memory in neocortex. Frontiers in neural circuits 10: 23.

Hawkins J., Subutai A., Yuwei Cui., 2017 A theory of how columns in the neo-cortex enable learning the structure of the world. Frontiers in neural circuits 11: 81.

Hessel M, Modayil J, Van Hasselt H, Schaul T, Ostrovski G, Dabney W, Horgan D, Piot B, Azar M, Silver D., 2018. Rainbow: Combining improvements in deep reinforcement learning. InProceedings of the AAAI conference on artificial intelligence 2018 Apr 29 (Vol. 32, No. 1).

Hohwy J., 2016. The self-evidencing brain. Noûs. Jun;50(2):259-85.

Huang Y, Rao RP, 2011. Predictive coding. Wiley Interdisciplinary Reviews: Cognitive Science. Sep;2(5):580-93.

Innocenti, G.M. and Price D.J., 2005. Exuberance in the development of cortical networks. Nature Reviews Neuroscience, 6 12, 955–965.

Klyubin AS, Polani D and Nehaniv CL., 2005. Empowerment: A universal agent-centric measure of control. IEEE Congress on Evolutionary Computation, volume 1, pp.128–135.

Laukien E., Crowder R., Fergal B., 2016. Feynman machine: The universal dynamical systems computer. arXiv preprint arXiv:1609.03971

Mnih V, Kavukcuoglu K, Silver D, Graves A, Antonoglou I, Wierstra D, Riedmiller M., 2013. Playing atari with deep reinforcement learning. arXiv preprint arXiv:1312.5602

Moran JM, Kelley WM, Heatherton TF., 2013. What can the organization of the brain's default mode network tell us about self-knowledge? Frontiers in human neuroscience. Jul 17;7:391.

Mountcastle VB, 1997. The columnar organization of the neocortex. Brain. Apr;120 ( Pt 4):701-22. doi: 10.1093/brain/120.4.701.

Nachum, Ofir, et al., 2018. Data-efficient hierarchical reinforcement learning. arXiv preprint arXiv:1805.08296

Pateria, Shubham, et al., 2021. Hierarchical Reinforcement Learning: A Comprehensive Survey. ACM Computing Surveys (CSUR) 54.5: 1-35.

Pivovarov I. and Shumsky S., 2022. Marti-4: new model of human brain, considering neocortex and basal ganglia–learns to play Atari game by reinforcement learning on a single CPU. International Conference on Artificial General Intelligence 2022, p. 62-74. doi:10.1007/978-3-031-19907-3_7

Rakic P., Bourgeois J.P., Eckenhoff M.F., Zecevic N., Goldman-Rakic P.S., 1986, Concurrent overproduction of synapses in diverse regions of the primate cerebral cortex. Science. Apr 11;232(4747):232-5. doi: 10.1126/science.3952506.

Seth A.K., Suzuki K., Critchley H.D.., 2012. An interoceptive predictive coding model of conscious presence. Frontiers in psychology. Jan 10;2:395.

Seth A.K., 2015. Presence, objecthood, and the phenomenology of predictive perception. Cognitive Neuroscience, 6(2–3), 111–117. doi:10.1080/17588928.2015.1026888

Seth A.K., Tsakiris M., 2018. Being a beast machine: The somatic basis of selfhood. Trends in cognitive sciences. 2018 Nov 1;22(11):969-81.

Shumsky, S.A., 2018. Deep structural learning: a new look at reinforcement learning. XX Russian Scientific Conference NEUROINFORMATICS 2018. Lectures on neuroinformatics, 11-43.

Shumsky, S.A., 2019. Machine intelligence. Essays on the theory of machine learning and artificial intelligence. RIOR Publishing, Moscow ISBN 978-5-369-02011-1.

Spratling M.W., 2017 A review of predictive coding algorithms. Brain and cognition 112: 92-97.

Stocco A, Lebiere C, Anderson J.R., 2010. Conditional routing of information to the cortex: a model of the basal ganglia's role in cognitive coordination. Psychological Review, 117 (2): 541–574. doi:10.1037/a0019077

Sutton, R.S., Doina P., Satinder S., 1999. Between MDPs and semi-MDPs: A framework for temporal abstraction in reinforcement learning. Artificial intelligence 112.1-2: 181-211.

Ungerleider L.G. and James V.H., 1994. 'What' and 'Where' in the Human Brain. Current Opinion in Neurobiology 4: 157–165

Ungerleider L.G. and Mishkin M., 1982. Two cortical visual systems. In D. J. Ingle, M. A. Goodale, & R. J. W. Mansfield (Eds.), Analysis of visual behavior (pp. 549-586). Cambridge: MIT Press.

Vezhnevets A.S. et al., 2017. Feudal networks for hierarchical reinforcement learning. International Conference on Machine Learning. PMLR.